\documentclass[useAMS,usenatbib]{mn2e}
\usepackage{amsmath,fleqn,graphicx,amssymb,aas_macros}
\usepackage{multirow}
\def\bolsig{\mbox{\boldmath$\sigma$}}
\def\bolx{\mathbf{x}}

\def\bole{\mathbf{e}}
\def\bolp{\mathbf{p}}
\def\hats{\mathbf{\hat{s}}}

\def\kpc{\,\mathrm{kpc}}
\def\kms{\,\mathrm{km\,s}^{-1}}

\def\bolv{\mathbf{v}}
\def\matA{\mbox{\boldmath{$\mathsf{A}$}}}
\def\matB{\mbox{\boldmath{$\mathsf{B}$}}}
\def\rsig{R_\sigma}
\def\pc{\,\mathrm{pc}}

\def\kpc{\,\mathrm{kpc}}
\def\pc{\,\mathrm{pc}}
\def\kms{\,\mathrm{km\,s}^{-1}}
\def\vsol{\mathbf{v_\odot}}

\title[Dangers of deprojection]
{The dangers of deprojection of proper motions}

\author[P.~J.~McMillan \& J.~J.~Binney]{
  Paul~J.~McMillan\thanks{E-mail: p.mcmillan1@physics.ox.ac.uk},
  and James~J.~Binney \\
  Rudolf Peierls Centre for Theoretical Physics, 1 Keble Road,
  Oxford, OX1 3NP, UK
}

\begin{document}
\maketitle

\begin{abstract}
  We re-examine the method of deprojection of proper motions, which
  has been used for finding the velocity ellipsoid of stars in the nearby
  Galaxy.   This method is only
  legitimate if the lines of sight to the individual stars are
  uncorrelated with the stars' velocities.  Very
  simple models are used to show that spurious results similar to ones
  recently reported are obtained when velocity dispersion decreases with
  galactocentric radius in the expected way. A scheme that compensates for
  this bias is proposed.
\end{abstract}


\section{Introduction}\label{sec:intro}

\citet[hereafter DB98]{DehnenBinney1998b} introduced a method for deprojecting
proper-motion data, which allowed them to explore the velocity distribution of
nearby stars in the Hipparcos catalogue~\citep{Hipparcos}, without
knowing their radial velocities. This works by taking a weighted ensemble
average of the proper motions of stars found in different parts of the
sky, under the assumption that the velocity distribution is
uncorrelated with position on the sky. This assumption was legitimate in the
case of the sample studied by DB98 because all its stars lay within
$\sim100\pc$ of the Sun,
so it was reasonable to approximate the full phase space distribution
function by the velocity-space distribution at the Sun: 
$f(\bolx,\bolv) \simeq f(\bolx_\odot,\bolv)$.

Recently \citet[hereafter F09]{Fuchsetal2009_short} used the DB98 technique to study a
sample of stars taken from the Sloan Digital Sky Survey \citep[SDSS:
][]{SDSS7_short}.  This data set contains stars that extend up to
$\sim800\pc$ above the plane and span a range of galactocentric radii $\sim2\kpc$
wide. Since the velocity dispersion of stars varies with both radius and
distance from the plane, the validity of the assumption that the velocity
distribution is uncorrelated with sky position is questionable for this
spatially extended sample.  In this paper we demonstrate that applying the
DB98 technique leads to erroneous results, particularly with regard to the
tilt of the velocity ellipsoid with respect to the Galactic plane. 

In Section~\ref{sec:depro} we briefly explain the DB98 method, and in
Section~\ref{sec:tests} we demonstrate that for a sample like that of F09 it
gives a biased estimate of the tilt of the velocity ellipsoid.
Section~\ref{sec:phys} explains the origin of this bias physically.
Section~\ref{sec:safe} proposes a technique for removing the bias. In
Section~\ref{sec:discuss} we discuss biases in the DB98 technique more
generally.

\section{Deprojection}\label{sec:depro}
The deprojection equations are stated and explained by
DB98, and written out in full by
F09. We repeat them here for clarity.

We work in a Cartesian coordinate system, centred on the Sun, in which
the $x$-axis points towards the Galactic centre, the $y$-axis points
in the direction of Galactic rotation, and the $z$-axis points towards
the north Galactic pole.  Given a star moving with heliocentric velocity $\bolv
\equiv (U,V,W)$, the observed proper-motion velocity is
\begin{equation} \label{eq:defp} \bolp = \bolv - v_\parallel\hats,
\end{equation}
where $\hats$ is the unit vector pointing from the
Sun to the star, and 
$v_\parallel$ is the component of $\bolv$ parallel to $\hats$. 
This can be written in matrix form as
\begin{equation} \label{eq:defpM}
 \bolp = \matA\cdot\bolv,\quad\mbox{where}\quad
  A_{ij} = \delta_{ij} - \hat s_i\hat s_j.
\end{equation}
The velocity ellipsoid is defined by both the mean velocity and the
velocity dispersion. To determine the velocity dispersion tensor we use the
equation
\begin{eqnarray}\label{eq:defsB}
  p_ip_j &=&  \sum_{k,l}A_{ik}v_k\,A_{jl}v_{l}\nonumber\\
&=&\sum_{kl}B_{ijkl}v_kv_l,
\end{eqnarray}
 where
\begin{equation}
B_{ijkl}\equiv\frac{1}{2}(A_{ik}\,A_{jl}+A_{il}\,A_{jk})
\end{equation}
 is the part  of $\matA\matA$ that is symmetric in its last pair
of indices.
 
We are interested in situations in which we know $\bolp$ and $\hats$ (and
therefore $\matA$ and $\matB$) but do not know $\bolv$.  It is clear from the
definition of $\bolp$ (equation~\ref{eq:defp}) that in this case we cannot
find $\bolv$ for an individual star because we do not know $v_\parallel$.
This is reflected in the fact that $\matA$ is singular.

We average equations (\ref{eq:defpM}) and (\ref{eq:defsB}) over a sample of
stars. If the velocities $\bolv$ of these stars are uncorrelated with their
sky positions $\hats$, they will be uncorrelated with $\matA$ and $\matB$,
and the expectation value of a product such as $\matA\cdot\bolv$ will equal
the expectation of $\matA$ times the expectation of $\bolv$. That
is, when the velocities are not correlated with $\hats$
 \begin{equation}
  \langle\bolp\rangle = \langle\matA\cdot\bolv\rangle = 
  \langle\matA\rangle \cdot \langle\bolv\rangle.
\end{equation}
 Provided the stars are sufficiently widely spread on the sky, the matrix
$\langle\matA\rangle$ is not singular, so we can write
 \begin{equation}\label{eq:findv}
  \langle\bolv\rangle = \langle\matA\rangle^{-1} \cdot \langle\bolp\rangle.
\end{equation}
 Similarly,
\begin{equation}\label{eq:findsig}
  \langle\mathbf{vv}\rangle = \langle\matB\rangle^{-1} \cdot 
  \langle\mathbf{pp}\rangle.
\end{equation}

\section{Tests}\label{sec:tests}

\begin{figure}
  \centerline{\resizebox{\hsize}{!}{\includegraphics{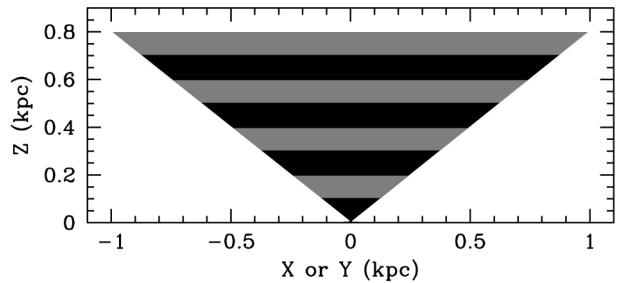}}}
  \caption{
Position of the 8 counting volumes, shaded black or grey alternately as we 
look further from the plane, that make up a cone with its vertex at 
the Sun, and its axis perpendicular to the Galactic plane.  
\label{fig:cone}
}
\end{figure}

In this section we demonstrate the danger of using the DB98 technique when
the key assumption of uncorrelated $\bolv$ and $\hats$ does not hold. We do
this by considering a simplified form of the problem addressed by F09, which
was to find the velocity ellipsoid of stars in the SDSS survey volume.
Fig.~\ref{fig:cone} shows our idealisation of the F09 counting volumes -- we
take them to be slices of a cone with the Sun as its apex.  In our usual
Cartesian coordinate system centred on the Sun, the cone is defined by
$\sqrt{X^2+Y^2}<0.8 Z$ and $Z<800\pc$. It is split into eight counting
volumes that are each $100\pc$ thick (cf Fig.~5 of F09).

F09 validated their use of the DB98 method by
drawing a velocity for every star in their sample from a Schwarzschild
distribution with a velocity ellipsoid that was everywhere aligned with the
$X,Y$ and $Z$ axes, and had constant axis lengths $\sigma_{U}$, $\sigma_{V}$
and $\sigma_{W}$.  This velocity distribution does not vary with position,
so $\bolv$ and $\hats$ will be uncorrelated. In reality the velocity
ellipsoid will vary from point to point, both in the orientation of its
principal axes, and in the lengths of these axes. The lengths of these axes
are expected to vary with galactocentric radius $R$ roughly as
$\bolsig\propto\exp(-R/R_\sigma)$, where $R_\sigma$ is of order twice the
disc's scale length $R_{\rm d}$ \citep[e.g.][]{GDII}.  In the Milky Way,
$R_{\rm d}\simeq2.5\kpc$ \citep[e.g.][]{Juricetal2008_short}, so
$R_\sigma\sim5\kpc$.

To illustrate the difficulty we adopt the distribution function
\begin{eqnarray}
  \label{eq:Schwarzcyl}
  f & \propto & \exp\left\{ -\frac{1}{2}\left[ \left( \frac{v_R}
        {\sigma_{R}(R,z)}\right)^2\right.\right.+\\
  & &\left.\left.\left( \frac{v_\phi-v_c(R)-
          \langle v_\phi(R,z)\rangle}
        {\sigma_{\phi}(R,z)}\right)^2+\left( \frac{v_z}
        {\sigma_{z}(R,z)}\right)^2\right]\right\}.
\end{eqnarray}
 where again $(R,\phi,z)$ are cylindrical coordinates centred on the Galactic Centre,
$v_c(R)$ is the circular speed, and $\langle
v_\phi(R,z)\rangle$ is the asymmetric drift.  The velocity
ellipsoid for this distribution function is aligned with the
cylindrical coordinate axes.
We assume that we can correctly compensate for the circular velocity
using Oort's constants \citep[e.g.][]{FeastWhitelock1997}. In all 
cases we take constant $\langle v_\phi(R,z)\rangle = -26\kms$. In
practice $\langle v_\phi\rangle$ varies with $\bolsig$, but this 
makes virtually no difference to these results, so we ignore it 
for simplicity.

We consider the following three forms for $\bolsig$:
\begin{enumerate}
\item \label{item:Fea} Constant $\bolsig
  \equiv  (\sigma_{R},\sigma_{\phi},\sigma_{z}) = (45,32,24)\kms$.
 This is nearly the same distribution function used by F09, except with the
velocity ellipsoid aligned with the cylindrical rather than Cartesian axes.

\vspace{3mm}

\item Radially varying  $\bolsig$
  \begin{equation}\label{eq:sigofR}
    \bolsig(R) = \bolsig(R_0)\,\exp[(R_0-R)/R_\sigma]   
  \end{equation}
 with $R_\sigma=5\kpc$, $R_0 = 8\kpc$, and $\bolsig(R_0)$ taking the
same value as in case~\ref{item:Fea}.

\vspace{3mm}

\item A form that varies both radially and vertically so as to provide
reasonable fits to the dispersions reported by F09:
  \begin{eqnarray}
  \bolsig(R,z) = (34 + 20z,23 &+&
  20z, 19 + 30z)\kms\nonumber\\
&\times&\exp[(R_0-R)/R_\sigma],
  \end{eqnarray}
  where $R_\sigma=5\kpc$ and $z$ is expressed in $\mathrm{kpc}$.
 \end{enumerate}

In each counting volume, we place 100,000 stars drawn randomly from a
uniform probability distribution over the entire volume. We assign
each star a velocity randomly chosen from the distribution
function. We then ``observe'' this star, and find its proper
motion. This allows us to compare the values of $\bolv$ and
$\bolv\bolv$ we determine from deprojection
(equations~\ref{eq:findv}~\&~\ref{eq:findsig}) to the real values.

Since we consider everything with respect to the Cartesian axes
defined in Section~\ref{sec:depro}, this yields values for $\langle U
\rangle$, $\langle UU \rangle$, $\langle UV \rangle$, etc. We can use
these values (and the fact that the centre of each counting volume lies
at $X=Y=0$) to find the velocity dispersions parallel to the cylindrical
axes, $\sigma_{R} 
,\,\sigma_{\phi}$ and $\sigma_{z}$ and the mixed moments $\sigma_{R\phi}^2
,\,\sigma_{Rz}^2$ and $ \sigma_{\phi z}^2$.

\begin{figure}
  \centerline{\resizebox{\hsize}{!}{\includegraphics{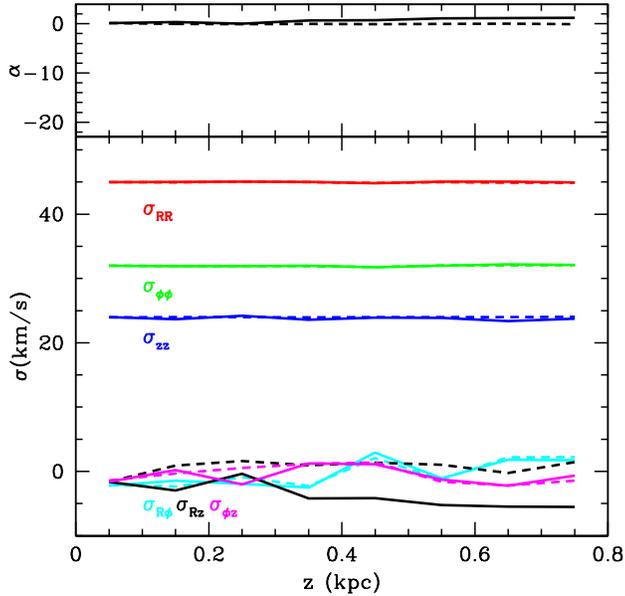}}}
  \caption{
Components of the velocity dispersion tensor $\bolsig$ (bottom) and the 
velocity ellipsoid tilt angle with respect to the plane, $\alpha$ (top) as a 
function of height above the plane. The figure shows the true values from 
the sample in each counting volume (dotted) and the values found by 
deprojection (solid). The true velocity ellipsoid has principal axes 
aligned with
the cylindrical coordinate directions and axis lengths that are independent
of position (i.e.~$\bolsig=\mathrm{const}$). 
\label{fig:varn}
}
\end{figure}

\begin{figure}
  \centerline{\resizebox{\hsize}{!}{\includegraphics{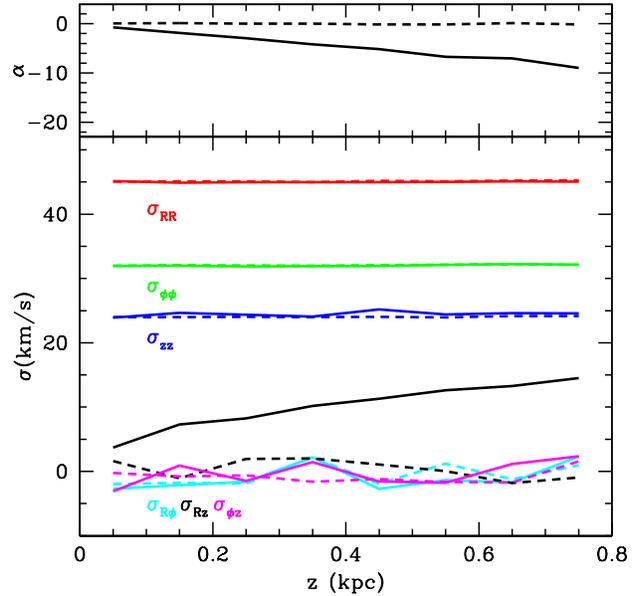}}}
  \caption{Similar to Fig.~\ref{fig:varn}, except for 
    $\bolsig\propto\exp(-R/R_\sigma)$, with $R_\sigma=5\kpc$. Again, 
    dotted lines show the true values, and solid lines show those found by
    deprojection.
\label{fig:varr}
}
\end{figure}

\begin{figure}
  \centerline{\resizebox{\hsize}{!}{\includegraphics{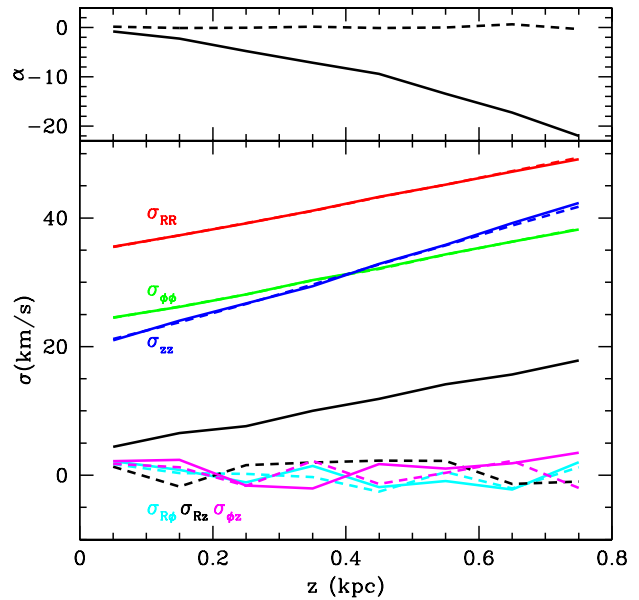}}}
  \caption{Similar to Figs.~\ref{fig:varn} and \ref{fig:varr}, with
    $\bolsig\propto\exp(-R/R_\sigma)$, with $R_\sigma=5\kpc$, and with
    $\bolsig$ varying with $z$ (as can be seen in the figure). Again, 
    dotted lines show the true values, and solid lines show those found by
    deprojection. The value of $\sigma_{Rz}$ found here is similar to 
    that found in Figure~\ref{fig:varr}, with the angle $\alpha$ being
    larger at large $z$ because the velocity ellipsoid is rounder, 
    due to $\sigma_{z}$ increasing more than $\sigma_{R}$ (c.f. 
    equation~\ref{eq:alpha}).
\label{fig:varzr}
}
\end{figure}

Note that the mixed moments may be either positive or negative.  In
Figs.~\ref{fig:varn}, \ref{fig:varr} \& \ref{fig:varzr} we plot
$\sigma_{R\phi}$, $\sigma_{Rz}$ and $\sigma_{\phi z}$, which we define
by
\begin{equation}
  \sigma_{ij}\equiv\mathrm{sign}(\sigma^2_{ij}) \sqrt{|\sigma^2_{ij}|}.
\end{equation} 
The two vertex deviations, which describe the orientation of the
velocity ellipsoid with respect to the cylindrical axes, can be found
from these values as
\begin{equation}
  \label{eq:psi}
  \Psi = -\frac{1}{2}\arctan{\frac{2\sigma^2_{R\phi}}
    {\sigma^2_{R}-\sigma^2_{\phi}}};
\end{equation}
\begin{equation}\label{eq:alpha}
  \alpha = -\frac{1}{2}\arctan{\frac{2\sigma^2_{Rz}}
    {\sigma^2_{R}-\sigma^2_{z}}}.
\end{equation}

In each of these cases, the values of $\langle U\rangle$, $\langle V\rangle$
and $\langle W\rangle$ determined from equation~(\ref{eq:findv}) are
consistent with the true values at $R_0$.

The lower panels of Figs.~\ref{fig:varn}, \ref{fig:varr} \& \ref{fig:varzr}
show the values of these velocity dispersions and the mixed moments as
functions of distance from the plane for the three distribution functions
described above: true values are shown by dotted lines, while solid lines
show values recovered by deprojection.  We see that deprojection yields
reasonably accurate values of $\sigma_{R}$, $\sigma_{\phi}$, $\sigma_{z}$,
$\sigma_{R\phi}$ and $\sigma_{\phi z}$ even when $\bolsig$ varies
significantly through the counting volumes, so the DB98 procedure is not
strictly valid.
 
However, the value of $\sigma_{Rz}$ found by deprojection is materially
incorrect in all cases, being slightly negative when $\bolsig$ does not vary
with $R$, and positive otherwise. The upper panels of
Figs.~\ref{fig:varn}, \ref{fig:varr} \& \ref{fig:varzr}
show that these incorrect values of $\sigma_{Rz}$ yield
values of the tilt angle as large as $\alpha\simeq-20\degr$.  A tilt of the
long axis of the ellipsoid towards the plane implied by $\alpha\simeq-20\degr$
is similar to that seen by F09. Thus our experiments demonstrate that the F09
tilt could be an artifact that arises because the velocity dispersion
increases inwards.

\subsection{Physical interpretation}\label{sec:phys}

To understand why a radial gradient in $\bolsig$ leads to an apparent tilt of the
velocity ellipsoid towards the plane, consider a simplified case in which
there are two
fields, both at Galactic coordinate $b=90-\theta$. One is at $l=0$ and the
other is at $l=180$. The velocity measured by the proper motion, $v_\mu$,
is then
\begin{equation}
  \label{eq:vmu}
  v_\mu =\left\{ 
    \begin{array}{ll}
      v_R\cos\theta+v_z\sin\theta,& \mbox{ at }l=0; \\
      v_R\cos\theta-v_z\sin\theta,& \mbox{ at }l=180.\\
    \end{array}
  \right.
\end{equation}
Since $0<\theta<90$, both $\sin\theta$ and $\cos\theta$ are positive.
Therefore, in the field at $l=0$, $v_\mu$ is large when $v_R$ and
$v_z$ have the same sign, while in the field at $l=180$ it is large
when they take opposite signs.  In the absence of a radial gradient,
the signature of a tilt \emph{towards} the plane is therefore larger
values of $v_\mu$ at $l=0$ than at $l=180$.  Clearly a radial gradient
in $\bolsig$ mimics this signature in the absence of a tilt. Hence if
one deprojects under the assumption that there is no radial gradient, the
algorithm will account for the data by reporting a tilt towards the plane.

\subsection{A workaround}\label{sec:safe}
 Given that good sky coverage is essential to the success of the DB09 method,
one simply cannot assume that the velocity distribution is the same at the
locations of all the stars in a sample that reaches out to $\ga1\kpc$ from
the Sun. A remedy that can be considered is to adopt a functional form for
the radial variation of $\bolsig$ and to use this form to correct the
observed proper-motion velocities to the values they would have had if
$\bolsig$ had been independent of position.  For example, for each star we
could calculate a ``corrected'' proper-motion velocity
 \begin{equation} \label{eq:correct} \bolp ' = (\bolp -\matA\cdot
  \mathbf{v}_{\mathrm{corr}}) \exp[(R-R_0)/R_\sigma ']
\end{equation}
 with $R_\sigma '$ an estimate for the true value of the parameter $R_\sigma$
that controls the radial variation of $\bolsig$ (eq.~\ref{eq:sigofR}) and
$\mathbf{v}_{\mathrm{corr}} = \vsol +\langle v_\phi\rangle\bole_\phi$ an
adjustment for the Solar motion and asymmetric drift. Thus defined, $\bolp'$
would be expected to average to zero over all directions and to be the
proper-motion velocity if there were no variation in $\bolsig$ with radius.

\begin{figure}
  \centerline{\resizebox{\hsize}{!}{\includegraphics{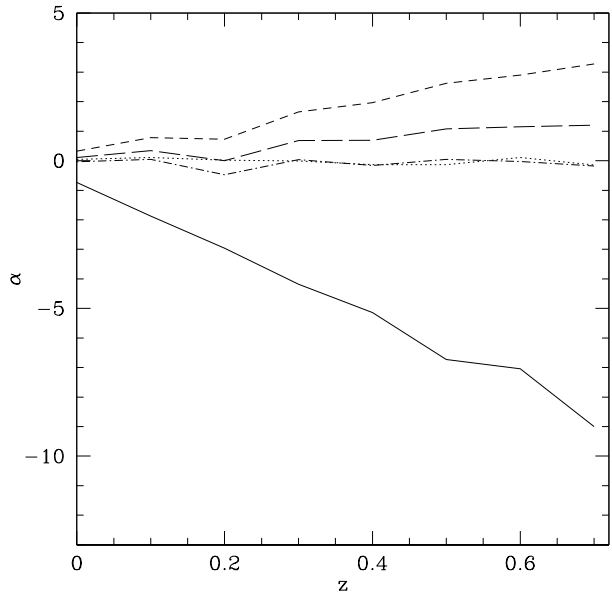}}}
  \caption{
 Tilt angle with respect to the Galactic 
 plane ($\alpha$) as a function of height above 
 the plane $z$. The true velocity ellipsoid is oriented parallel to the
 cylindrical axes, with $\bolsig\propto\exp(-R/(5\kpc))$. The figure shows 
 the true value of $\alpha$ for the sample (dotted), the value 
 found from the proper motions without applying any correction 
 (solid), and values found applying a correction 
 (equation~\ref{eq:correct}) with $R_\sigma '= 4\kpc$ (short dashed), 
 $5\kpc$ (long dashed) and $6\kpc$ (dot dashed). The ``true'' 
 correction, $R_\sigma '=5\kpc$ does not return the true value 
 of $\alpha$ because it does not correct for the fact that the 
 velocity ellipsoid is not aligned with the Cartesian axes.
\label{fig:corr}
}
\end{figure}

We test this correction by applying it to simulated data generated as in case (ii)
above.  We know the true value of
$\bolv_{\mathrm{corr}}$ in this case, so we ignore the relatively
minor uncertainties which are caused by not estimating this correctly.

The dashed lines in Fig.~\ref{fig:corr} show the tilt angle $\alpha$ found
from corrected proper-motion velocities for three values of $R_\sigma '$:
$4\kpc$ (short-dashed), $5\kpc$ (long-dashed) and $6\kpc$ (dot-dashed). In
all three cases the corrected data give much more accurate results than the
uncorrected data (full curve), but the most accurate results are obtained
with $R_\sigma'=6\kpc$ rather than the true value, $5\kpc$; with $R_\sigma '
= 5\kpc$ we find $\alpha\sim1\degr$ at the largest values of $z$ because the
correction does not address the problem that the axes of the velocity ellipsoid
are aligned with the cylindrical rather than Cartesian axes. A closely
related bias is seen when $\bolsig$ is constant (Fig.~\ref{fig:varn}).
Using a value of $R_\sigma ' = 6\kpc$ for the correction gives
$\alpha\simeq0$ because it \emph{under}-compensates for the bias due to the
variation in $\bolsig$, which inadvertently compensates for the bias due to
the alignment of the velocity ellipsoid's axes.

If we considered the value $\alpha$ well established, we could use corrected
proper-motion velocities to determine $R_\sigma$ from the data.

\section{Discussion}\label{sec:discuss}
 In this paper we have focused on the tilt of the velocity ellipsoid towards
the plane, and may have left the reader with the impression that, for
example, the tilt in the plane or the non-mixed terms ($\sigma_{R}$ etc.) are
correctly recovered by the DB98 technique. While this is true to a good
approximation in the cases  shown here, it is not always true.

For example, consider the situation described in Section~\ref{sec:safe},
in which we need to know the value of $R_\sigma$ so  we can compensate for
the variation in $\bolsig$ across the counting volume. In an approach to the
determination of  $R_\sigma$ we might split the data into two
sets, for $|l|<90\deg$ and $|l>90\deg$, and find $\bolsig$
separately for each set -- this gives us enough information to find $\rsig$.
However, if the data are split in this way, they produce a bias in the values
of the \emph{non}-mixed components of $\bolsig$ (as well as the mixed
components). This bias is in opposite directions for the two
data sets, so strongly affects the derived value of $\rsig$, but cancels out
when the two sets are considered together (hence the lack of bias in the
non-mixed components in Figs.~\ref{fig:varr}~and~\ref{fig:varzr}).

Similar biases must \emph{always}
be considered when using deprojection. In the tests described above,
the symmetry of the counting volumes cancelled out the bias in most
components of $\bolsig$, effectively restricting it to $\sigma_{Rz}$. The counting
volumes of real data sets will not enjoy the high degree of symmetry
characteristic of our model sets, with the result that biases in the values
returned by the DB98 method will not be confined to
$\sigma_{Rz}$. 

\section{Conclusions}

In this paper we have demonstrated that the statistical deprojection of
proper motions cannot be applied straightforwardly to data spanning a
significant
volume of the Galaxy.  This is primarily because the dependence of the
velocity dispersion $\bolsig$ on position violates the central assumption of
the method.

Using a simple model we have demonstrated that applying this method can
suggest a large tilt of the velocity ellipsoid towards the plane, even if the
actual tilt is zero. It seems very likely that this effect is responsible for
the remarkably large tilt, $\alpha=-20\degr$, reported by F09. Correcting for
this effect in the manner discussed in Section~\ref{sec:safe} would probably
bring this result much closer to the smaller tilt angles obtained using
radial velocities \citep[e.g.][]{Siebertetal2008_short,Bondetal2009_short}.

We note, however, that all components of $\bolsig$ other than $\sigma_{Rz}$
were nearly unaffected by this bias in our tests. For a realistic survey
volume such as that used by F09, these biases are likely to be larger than in
our tests and in some circumstances may materially affect the results.

\section*{Acknowledgments}

This research was supported by a
grant from the Science and Technology Facilities Council.

\bibliographystyle{mn2e} \bibliography{refs}
\end{document}